\documentclass[twocolumn,secnumarabic,amssymb, nobibnotes, prb]{revtex4}
\usepackage{bm}%
\usepackage{graphicx}
\usepackage{subfigure}
\usepackage{multirow}
\usepackage{color}

\begin{document}
\title{Tuning the caloric response of BaTiO$_3$ by tensile epitaxial strain}

\author{Anna Gr\"unebohm}
\email{anna@thp.uni-due.de}
\affiliation{Faculty of Physics and CENIDE, University of Duisburg-Essen, 47048 Duisburg, Germany}
\author{
Madhura Marathe}
\author{
Claude Ederer}
\affiliation{Materials Theory, ETH Z\"urich, 8093 Z\"urich, Switzerland}

\begin{abstract} We investigate the effect of epitaxial strain on the
  electrocaloric effect (ECE) in BaTiO$_3$ by means of {\it{ab
      initio}} based molecular dynamics simulations.  We show that
  tensile strain can be used to optimize the operation range for
  ferroic cooling.  Strain in the range of $\leq 1$\,\% can be used to
  shift the operation temperature by several hundreds of Kelvin both
  to higher and lower temperatures, depending on the direction of the
  external field. In addition, the transformation between multi-domain
  and mono-domain states, induced by an in-plane electric field,
  results in an additional peak of the adiabatic temperature change at
  lower temperatures, and a broad temperature interval where the
  caloric response scales linearly with the applied field strength,
  even up to very high fields.
  \end{abstract}

\maketitle


\section{Introduction}
Within a ferroic material, a variation of the ferroic order parameter
can be induced through application of the corresponding conjugate
field. Depending on the thermal boundary conditions, this results in
an isothermal entropy or an adiabatic temperature change. For example,
if an externally applied conjugate field is removed under adiabatic
conditions, order parameter and temperature typically decrease.
It has been shown that ``giant'' caloric effects occur for different
classes of materials in the vicinity of field-induced 
structural-ferroic transitions~\cite{Planes,Moya}. For example, in ferroelectric BaTiO$_3$ (BTO), the largest
field-induced change of the order parameter (electric polarization,
$P$), and thus the largest adiabatic temperature change, is observed
for temperatures slightly above the transition between the cubic
paraelectric high temperature phase and the tetragonal ferroelectric
phase at about 400~K \cite{Bai,Moya2}. 
Systems that exhibit such giant caloric effects have a huge potential
for the development of novel cooling concepts
\cite{ferrocooling,Valant,Takeuchi/Sandeman:2015}.
For the case of the electro-caloric effect
(ECE)~\cite{Scott:2011,Valant}, much work is focused on thin films \cite{Mischenko_et_al:2006,Bai,Moya,Scott:2011,Valant},
as the magnitude of the ECE generally increases
with the strength of the applied electric field~\cite{Lisenkov}, and
large fields can be induced by moderate voltages within thin
films~\cite{Mischenko_et_al:2006}. In addition, the high surface to
volume ratio allows for a fast heat transfer to the environment and
thus, potentially, for fast cycling of a device. Thin films also allow
to combine the advantages of single crystals (well defined
crystallographic orientation) and poly-crystals (mechanical stability
during cycling). 

Furthermore, thin films are often subject to strain, which can
strongly affect the ferroelectric properties, and thus provides an
efficient way for tailoring the desired functional
properties \cite{Schlom}.
\begin{figure}
\centering{ \includegraphics[height=0.35\textwidth,clip,trim=2cm 0.5cm
    4cm 3cm]{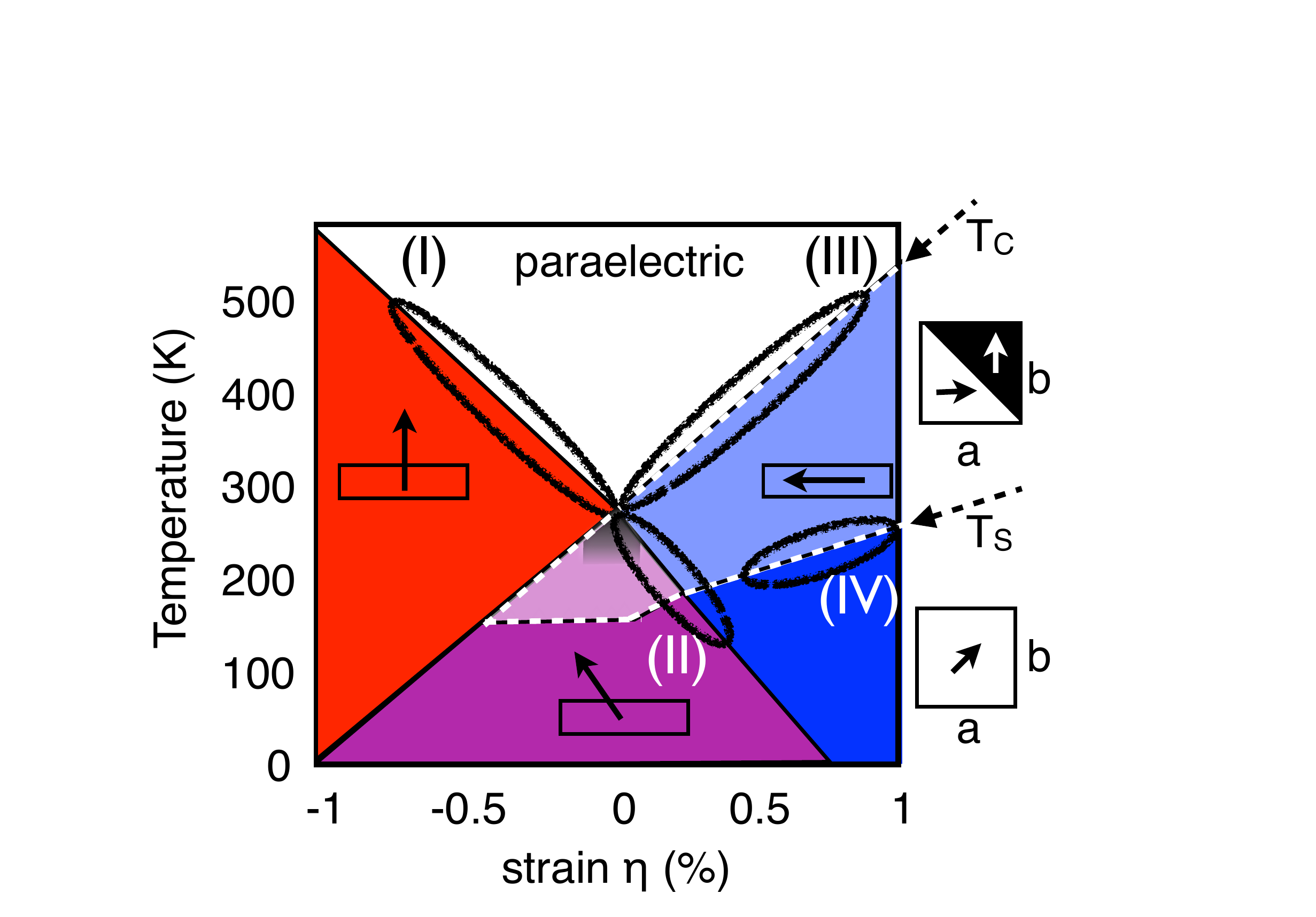}
\caption{Schematic phase diagram of BaTiO$_3$ under epitaxial strain,
  adapted from \cite{BTO_strain}. Colors encode different polarization ($P$)
  directions (also illustrated by the insets). Red: $P$ along $c$;
  light blue: $a$/$b$ multi-domain state ($P$ as sketched on the upper
  right-hand side); darker blue: $ab$ mono-domain state ($P$ as sketched
  on the lower right-hand side); light purple: $ac$/$bc$ domains (in plane $P$ as for $a/b$ domains with homogenous $P$ along $c$);
  darker purple: $abc$ mono-domain phase. White dotted lines border
  multi-domain phases. $T_{\text{C}}$ ($T_{\text{S}}$) indicates the strain-dependent
  transition temperature for in-plane $P$ (for the
  multi-domain to mono-domain transition).
The grey area below the bulk $T_{\text{C}}$  for $\eta\approx 0$ indicates a region with complex multi domain states.
\label{fig:phase}}}
\end{figure}
If a material such as BTO is grown epitaxially on a surface
representing a square lattice, a biaxial strain, $\eta$, is imposed
along the two in-plane directions ($a$ and $b$),
whereas the lattice constant along the film normal ($c$) can adjust
freely. As a result, the ferroelectric transition temperatures and the
relative stability of different ferroelectric phases of BTO are
modified, as depicted in
fig.~\ref{fig:phase}~\cite{pertsev2,pertsev4,pertsev3,Dieguez,Li2,BTO_strain}.
For zero strain, {\it{i.e.}} pure clamping by the substrate, both in-plane
and out-of-plane components of the polarization appear at the same
transition temperature $T_{\text{C}}$, equal to the bulk transition
temperature. Tensile (positive) strain then increases the transition
temperature for the in-plane components of $P$ ({\it{i.e.}}, along $a$ and/or
$b$) and decreases the transition temperature for the out-of-plane
component (along $c$), and vice versa for compressive
strain~\cite{pertsev2,Dieguez}.  In addition to the ferroelectric
mono-domain phases found at low temperatures, multi-domain
configurations are stabilized between $T_{\text{C}}$ and
$T_{\text{S}}$~\cite{pertsev3,Li2,Qiao,BTO_strain,Everhardt}.

Generally, a field-induced change of polarization and a
resulting adiabatic temperature change occur for all values of
strain and all temperatures. However, the largest effect is usually observed just above $T_{\text{C}}$. This can be understood from the following relation (cf.~ref. \cite{Valant}):
\begin{equation}
\Delta T = - \int_{E_{1}}^{E_{2}} \frac{T}{C_E(T,E)}
\left. \frac{\partial P}{\partial T}\right|_E dE \quad.
\label{eq:ECE}
\end{equation}
Here, the electric field $E$ is varied from $E_1$ to $E_2$ under
adiabatic conditions, $P$ is the polarization component along the
field direction, and $C_E$ is the specific heat at constant field (and
constant pressure). The largest changes, $\partial P/\partial T<0$,
occur just above $T_{\text{C}}$ (where the field causes a transition
or crossover to the ferroelectric phase), resulting in a strong peak
of $\Delta T(T)$ with a relatively sharp drop on the low temperature
side and a broad shoulder towards higher temperatures, see
{\it{e.g.}} refs.~\cite{Moya,Valant}.

Most previous studies investigating the influence of epitaxial strain
on the ECE in BTO have concentrated on compressive strain
\cite{Cao,Zhang3,Madhura,Akcay}. In contrast, the influence of tensile
strain has so far only been investigated based on phenomenological
models, with partly contradicting results. For example, both an
increasing~\cite{Akcay} and a decreasing~\cite{Cao,Zhang3} maximal
temperature change have been reported. Such discrepancies, which can
be caused by different model assumptions or parameterizations, can be
resolved through {\it ab initio}-based simulations.  Furthermore, for
both kinds of strain we are only aware of investigations on the role
of an electric field along the surface normal. However, the ECE in
single crystals can depend strongly on the direction of the applied
field \cite{Perantie,Ponomareva,LeGoupil_et_al:2014}. In addition, the
ECE in the multi-domain states with $a/b$ and $ac/bc$ domains and in
particular at the corresponding multi-domain to mono-domain
transitions at $T_{\text{S}}$, has not been investigated so far.

In this paper, we address the following questions: Does tensile strain
enable or hamper a large caloric response? In particular, can such
strain help to obtain a significant ECE over a broad temperature range
around ambient temperatures, which is desirable for many applications?
How does the strain-induced multi-domain state affect the ECE?  To
understand these issues, we explore the role of tensile strain on the
caloric response of BTO, using {\it ab initio}-based simulations and
taking different directions of the applied field into account.  We
show that tensile strain $\leq 1$\,\% can be used to shift
$T_{\text{max}}$ by several hundreds of Kelvin both to higher and lower
temperatures, depending on the direction of the external field.  In
addition, the transformation between multi-domain and mono-domain
states, induced by an in-plane electric field, results in an
additional peak of $\Delta T(T)$ at lower temperatures, and a broad
temperature interval where the caloric response scales linearly with
the applied field strength, even up to very high fields.


\section{Methods}
\label{sec:comp}

Molecular dynamics simulations are performed, employing an effective
Hamiltonian \cite{Zhong} as implemented in the {\sc feram} code
developed by Nishimatsu {\it{et al.}}
\cite{Feram1}.\footnote{\texttt{http://loto.sourceforge.net/feram/}}
The {\it{ab initio}}-based parametrization of the effective
Hamiltonian for BTO is taken from ref.~\cite{Nishimatsu}. Periodic
boundary conditions and a cell size of 96$\times$96$\times$96 BTO
units have been used. We model the epitaxial strain imposed through a
hypothetical substrate by fixing the elements $\eta_1$ ($\equiv
\eta_a$) and $\eta_2$ ($\equiv \eta_b$) of the homogenous strain
tensor (in standard Voigt notation) to the external strain $\eta$ and
setting $\eta_6=0$.  We define zero strain for a lattice constant of
3.996~{\AA}, which we obtain for the paraelectric phase of bulk BTO
directly above $T_{\text{C}}$.  

We note that, due to the large size of our simulation cell, we obtain
additional domain configurations, with local polarization along
various crystallographic directions, for very small strain values
($|\eta|<0.1$\%) directly below $T_C$ (see grey area in
fig.~\ref{fig:phase}). These configurations are absent for the smaller
simulation cell used in ref.~\cite{BTO_strain}.  Since these
additional domain configurations appear only in a very limited region
within the phase diagram they are not relevant in the context of the
present study and are therefore not discussed any further.

The caloric response is obtained from the following protocol: (I)~We
start our simulations above $T_{\text{C}}$ and equilibrate the system
within an external field. Afterwards, we reduce the temperature in
steps of maximal 10~K, using the final configuration from the previous
temperature step as initial configuration, {\it{i.e}}.\ we field-cool the
system. These simulations are performed in the $NPT$ ensemble using a
Nos{\'e}-Poincar\'e thermostat and a time step of $\Delta t = 2$~fs
\cite{Nose}. (II)~At each measuring temperature we start from the
corresponding field-cooled configuration, switch to the
micro-canonical ($NPE$) ensemble and perform \emph{direct} simulations
of the ECE. 
Advantages and technical details of this method are discussed in refs.~\cite{Nishimatsu3,Marathe,Liu2}.  We ramp down the field with
the rate of 0.002~kV/cm/fs. This ramping rate is sufficiently slow to
ensure thermodynamic equilibrium except very close to the
ferroelectric phase transition, where the dynamic of the system slows
down considerably.  We thus use further equilibration of at least
40~ps previous to measuring the adiabatic temperature change.

We note that the simplifications necessary to construct the effective
Hamiltonian lead to quantitative deviations of the calculated
ferroelectric transition temperatures ($T_{\text{C}}$) compared to
experiment~\cite{Zhong,Nishimatsu}.
Nevertheless, qualitative trends and the overall magnitude of polarization and strain are well
described and are generally in good agreement with experimental
observations, cf.~refs.~\cite{Dieguez,BTO_strain}.
 Furthermore, only 3 out of of 15
degrees of freedom per BTO unit are explicitly taken into account in
our simulations, resulting in a reduced specific heat, and thus an
overestimated temperature change $\Delta T$. In the following, we
therefore rescale the calculated $\Delta T$ by a factor of $1/5$,
according to the ratio between the considered and total number of
degrees of freedom, cf.~refs.~\cite{Nishimatsu3,Defects}.

 
\section{Results}
\begin{figure}
\includegraphics[height=0.35\textwidth,clip,trim=3.2cm 2cm 2cm 2cm]{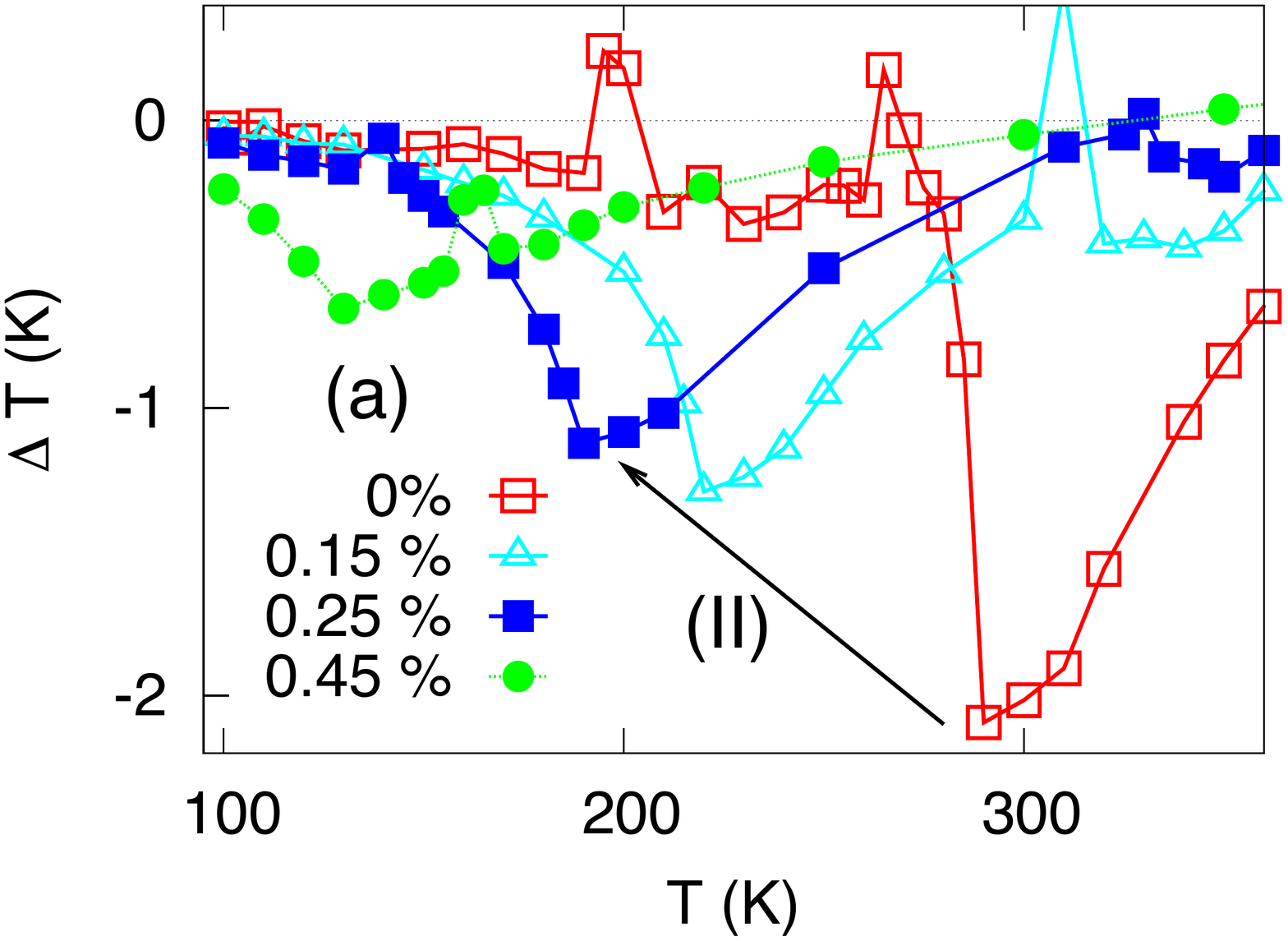}
\includegraphics[height=0.35\textwidth,clip,trim=3.2cm 2cm 2cm 2cm]{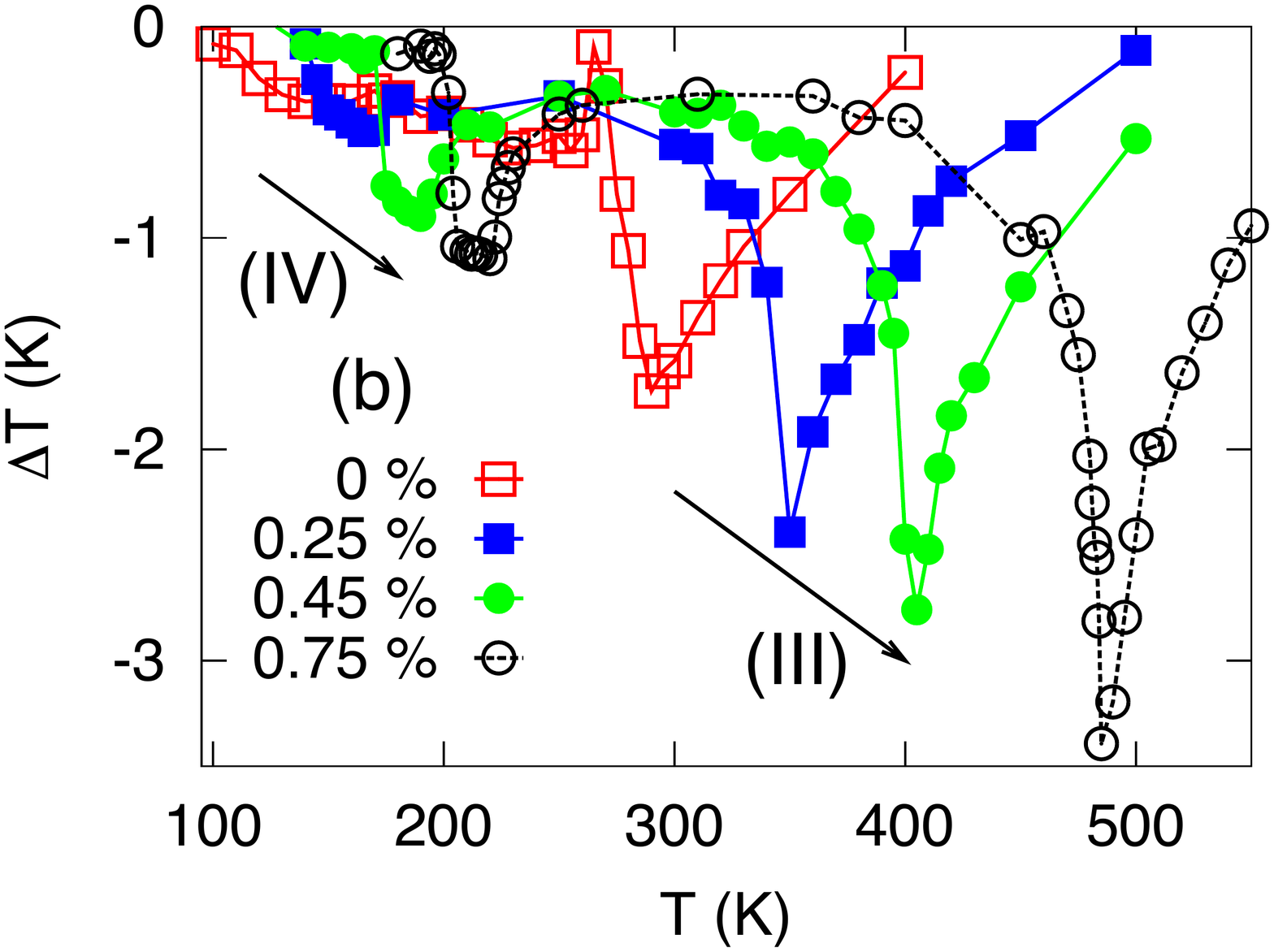}
\caption{Adiabatic temperature change of BTO if an external field of
  100~kV/cm is removed for different values of tensile strain. a):
  Field along $c$; b): field along $ab$. Arrows illustrate the
  changes of $T_{\text{max}}$ and $\Delta T(T_{\text{max}})$ with strain
  in the vicinity of the three transition lines marked (II)-(IV) in
  fig.~\ref{fig:phase}. 
\label{fig:ece}}
\end{figure}
Figure~\ref{fig:ece} shows the calculated electro-caloric temperature
change as function of the initial temperature under removal of an
electric field of 100 kV/cm applied along the out-of-plane and
in-plane directions for different amounts of tensile
strain. 

We first discuss the case of zero applied strain, $\eta=0$, {\it{i.e.}}, pure
clamping, for both field directions. As already discussed in
refs.~\cite{Madhura,Akcay} clamping reduces the ECE and results in a
broadening of the $\Delta T$-peak.  For the field along the surface
normal we find a reduction by a factor of two compared to the free
bulk system (from about 4 to 2\,K for an external field of 100\,kV/cm)
and the full width at half maximum (FWHM) increases from about 35~K
(unclamped) to about 55~K (clamped).  If the field is applied along
the in-plane $ab$ direction, the peak value of the ECE, $\Delta
T(T_{\text{max}})$, is reduced to about 1.7 K at $T_{\text{max}}=290$~K,
while the FWHM increases to about 70~K.  Thus, in spite of the
reduction compared to the free bulk case, a large ECE is still
obtained for the clamped system for both field directions.

We note that we find small temperature and field regions where
$\partial P/\partial T |_E>0$, {\it{e.g.}} at the transition temperatures of
the ferroelectric phase polarized perpendicular to the external field.
As a consequence, the adiabatic cooling in these temperature regions
is reduced and even an inverse ECE, {\it{i.e.}}, a heating of the system
under field removal occurs, cf.\ eq.~(\ref{eq:ECE}).  For instance, at
the transition temperature of the in-plane polarization, $T_C \approx
270$\,K, for field along $c$, $\Delta T$ changes its sign.  In
addition, an inverse ($\Delta T>0$) peak exists around $T \approx
190$\,K, the transition temperature to a different multi-domain
configuration observed for $\eta\approx 0$.  The magnitude of these
inverse peaks may be underestimated due to the used temperature grid
and as several (meta-) stable domain configurations seem to
coexist.

Focusing next on the strain dependence of $\Delta T(T)$, depicted in
fig.~\ref{fig:ece}, one can observe that increasing tensile strain
shifts the position of the peak in $\Delta T$, $T_{\text{max}}$,
towards higher (lower) temperatures if the electric field is applied
in-plane (out-of-plane). Thus, the peak positions essentially follow
the transition lines (II) and (III) in fig.~\ref{fig:phase} for the
appearance of spontaneous polarization along the respective field
direction. Furthermore, with decreasing (increasing) $T_{\text{max}}$,
the FWHM of the ECE peak increases (decreases).  For example, for a
field along $c$ the FWHM increases approximately linearly with strain
from about 55~K to about 90~K between 0 and 0.45~\% while the FWHM for
a field along $ab$ is about 50~K for 0.25--0.45~\% and is reduced to
40~K for 0.75\% strain.

Most strikingly, the maximum of the ECE,  $|\Delta
T(T_{\text{max}})|$, increases (decreases) systematically, if the
corresponding peak position is shifted to higher (lower) temperatures.
This relationship between $|\Delta T(T_{\text{max}})|$ and
$T_{\text{max}}$ for different strains can be understood from
eq.~\ref{eq:ECE}.  Even though all quantities under the integral in
eq.~\ref{eq:ECE} depend on temperature (and of course on the
electric field), we find that the change of the peak value of $\Delta
T$ under strain is dominated by the explicit factor $T$ in the
integrand.
This is analogous to what has been found in ref.~\cite{Madhura} for
BTO under compressive strain and electric field along $c$,
see the corresponding transition line (I) in fig.~\ref{fig:phase},
using the same microscopic model as used in the present work.

The same trend, {\it{i.e}}.\ increasing $|\Delta T (T_{\text{max}})|$ with
increasing $T_{\text{max}}$, due to the strain-induced shift in the
ferroelectric transition temperature, has also been found in
\cite{Cao} using a modified transverse Ising model and in
\cite{Zhang3} using phenomenological Landau theory.
In contrast, ref.~\cite{Akcay}, also using Landau theory, has reported
the opposite trend, namely decreasing $|\Delta T(T_{\text{max}})|$ with
increasing $T_{\text{max}}$ (and thus increasing $|\Delta
T(T_{\text{max}})|$ under increasing tensile strain). The increase of
$|\Delta T|$ found in ref.~\cite{Akcay} was attributed to a stronger
first order character, and thus increasing $\partial P/\partial T
|_E$, under tensile strain, cf.\ eq.~\ref{eq:ECE}. This can neither
be confirmed by our {\it ab initio}-based microscopic model nor by the
phenomenological models used in \cite{Zhang3,Cao}.  We note that,
while the Landau free energy in \cite{Akcay} is expanded up to sixth
order in the polarization, with temperature-dependent second and
fourth order coefficients, ref.~\cite{Zhang3} uses a different
parameterization based on an eighth-order expansion with a
temperature-independent fourth order coefficient.

In addition to the main peak at $T_{\text{max}}$, corresponding to the
ferroelectric transition temperature of the polarization component
along the field direction, for small values of tensile strain and
field along $c$ we also observe a small feature in $\Delta T$ at the
in-plane transition temperature $T_C^{ab}> 300$\,K. Here, the onset of
the in-plane polarization results in a reduced magnitude (or even
reversed sign) of $\Delta T$, related to regions with negative
$\partial P_c/\partial T |_E$. We note that this is analogous to the
``double-peak structure'' reported in \cite{Zhang3} for tensile strains
smaller than 0.15\,\%. With increasing strain, the magnitude
of the ``peak'' at higher temperatures decreases as $T_C^c$ and $T_C^{ab}$
move apart, see fig.~\ref{fig:phase}.

Summarizing the results presented so far, we point out that epitaxial
strain can be used to systematically increase or decrease the
temperature range with large ECE over several hundreds of Kelvin for
rather moderate strain values of less than 1\,\%, depending also on
the direction of the applied field. In fact, for many of the
anticipated device applications, it is important to reduce
$T_{\text{max}}$ to ambient temperatures (and also to obtain a
sufficiently large $\Delta T$ over a broad temperature interval around
$T_{\text{max}}$). Here, we find $T_{\text{max}}$ about 100~K below the
bulk $T_{\text{C}}$, and thus in a very attractive temperature range
for applications, for only 0.25\% tensile strain (see
fig.~\ref{fig:ece}a).

Next, we discuss the ECE for the multi-domain state observed under
tensile strain in the temperature range between $T_{\text{C}}$ and $T_{\text{S}}$.
As can be seen in fig.~\ref{fig:ece}b for an electric field applied
along $ab$, a second peak in $\Delta T(T)$ is found near the
strain-dependent transition temperature $T_{\text{S}}$, indicated as (IV) in
fig.~\ref{fig:ece}.
The appearance of this peak can be explained as follows. Below $T_{\text{S}}$,
the system is in a mono-domain state with a large polarization that
depends only weakly on temperature and field. Consequently the system
exhibits only a very weak ECE below $T_{\text{S}}$. If an external field along
$ab$ is applied, $T_{\text{S}}$ shifts to higher temperatures, {\it{e.g.}}, for 0.75\%
strain, $T_{\text{S}}$ increases from 185~K (no field) to 220~K (100~kV/cm) to
370~K (500~kV/cm). Therefore, if the external field is increased
starting from an initial temperature slightly above $T_{\text{S}}$, the system
transforms reversibly from a multi-domain to a mono-domain
state. Since the net polarization along $ab$ in the multi-domain
$a$/$b$ state is smaller than in the mono-domain $ab$ phase, the
polarization is strongly field- and temperature-dependent around this
transition, resulting in a pronounced peak in $\Delta T$.

While for $\eta=0$ the corresponding peak value $\Delta
T(T_{\text{max}}^S)$ for an external field of 100~kV/cm is less than
0.5~K, $\Delta T(T_{\text{max}}^S)$ increases with increasing tensile
strain and becomes larger than 1\,K for 0.75\,\% strain at about 75~K
below the zero-strain $T_{\text{C}}$, see fig.~\ref{fig:ece}~(b).   This increase of $\Delta
T(T_{\text{max}}^S)$ is again related to the corresponding increase of
$T_{\text{max}}^S$ via eq.~\ref{eq:ECE}, as discussed above for the
main ECE peak.

Apart from the additional peak in $\Delta T$, we also observe a
moderately large ECE ($\sim$ 0.5~K for 0.75~\% strain) within a rather
broad temperature interval between the two main ECE peaks,
namely in the temperature range between $T_{\text{S}}$ and $T_{\text{C}}$ corresponding to the strain-induced multi-domain configuration. While the corresponding $\Delta T$ is small
compared to the peak values, it is noticeably larger than within the
ferroelectric mono-domain phase below $T_{\text{S}}$. This enhanced ECE is due
to the field-induced (and temperature-dependent) gradual rotation of
the local polarization from $a/b$ towards $ab$.
Furthermore, as we show in the following, the ECE in this multi-domain
region can be considerably enhanced by increasing the field strength.

\begin{figure}
\includegraphics[width=0.45\textwidth,clip,trim=3cm 5.4cm 1cm 2cm]{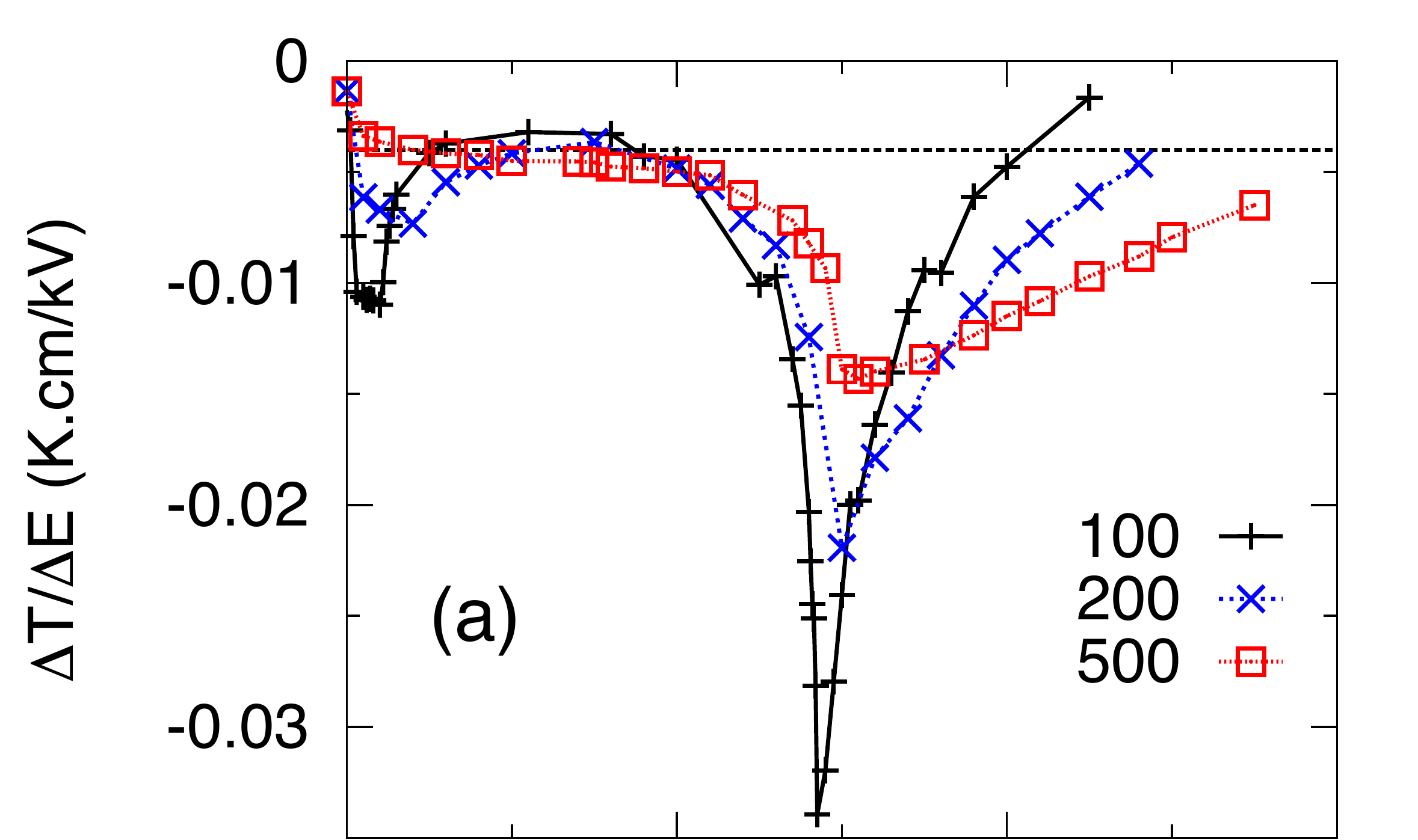}
\includegraphics[width=0.45\textwidth,clip,trim=3cm 2cm 1cm 2.4cm]{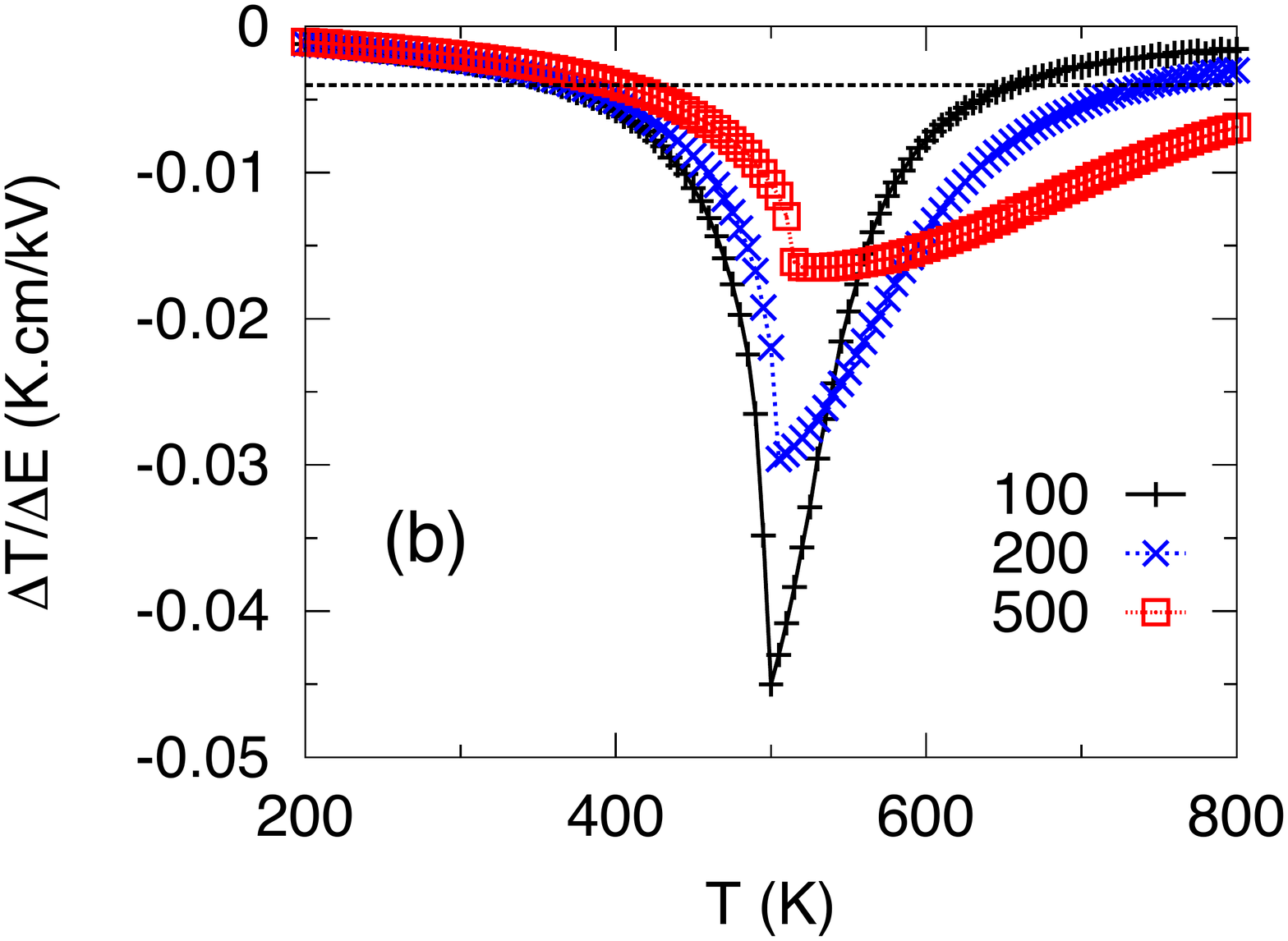}
\caption{Electro-caloric strength $\Delta T/\Delta E$ for BTO under
  $\pm$0.75\,\% tensile (a) and compressive (b) strain for electric
  fields applied along $ab$ (a) and along $c$ (b). Dotted horizontal
  lines mark an electro-caloric strength of $-$0.004\,K$\cdot$cm/kV.
\label{fig:ece_strength}}
\end{figure}

Figure~\ref{fig:ece_strength} shows the ``electro-caloric strength'',
defined as $\Delta T/\Delta E$ \cite{Moya}, both under compressive and
tensile strain for different applied fields. It can be seen that in
both cases the electro-caloric strength decreases near $T_{\text{max}}$
while it increases for even higher temperatures as the field
increases.  For lower temperatures, which are most relevant for
practical cooling applications, the electro-caloric strength under
compressive strain is very small and decreases further with the field
strength.  In contrast, the electro-caloric strength under tensile
strain is approximately constant between $T_{\text{C}}$ and
$T_{\text{S}}$, around $-$0.004\,K$\cdot$cm/kV, even for field
strengths up to 500~kV/cm. Thus, between $T_{\text{C}}$ and
$T_{\text{S}}$ the ECE grows linearly with the field, and already for
a field strength of 200~kV/cm an adiabatic response of 1~K can be
obtained in a very broad temperature interval below $T_{\text{C}}$.
For 0.75\% strain, this large and broad response occurs about 100~K
below the bulk $T_{\text{C}}$, {\it{i.e}}.\ corresponding to ambient
temperatures.

\begin{figure}
\includegraphics[width=0.45\textwidth,clip,trim=3cm 2cm 1cm 2cm]{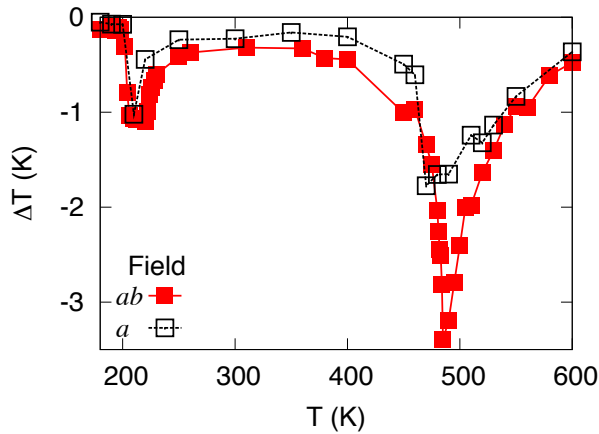}
\caption{Adiabatic temperature change of BTO at 0.75\% tensile strain
  under removal of an external field of 100~kV/cm along the $a$ (open
  black squares) and $ab$ (filled red squares) directions.
\label{fig:ece_100}}
\end{figure}

Up to now, the in-plane electric field in our calculations was always
oriented along the $ab$ direction. However, we note that even for a
misaligned field, a rather strong ECE near $T_{\text{S}}$ is
obtained. For example, for an in-plane field along $a$, we find a
similar $\Delta T(T)$ profile and a similar ECE peak at $T_{\text{S}}$
as for the case with the field along $ab$, see fig.~\ref{fig:ece_100}.
Only the magnitude of the ECE between $T_{\text{C}}$ and
$T_{\text{S}}$ and close to the peak at $T_{\text{C}}$ are reduced.
This reduced $\Delta T$ for $E$ along $a$ is most likely due to the
polarization-strain coupling, which does not allow the polarization to
rotate fully into the field direction, due to the epitaxial
constraint.  

Finally, we note that we use ``poled'' (field-cooled) starting
configurations in our simulations, and therefore our results do not
include irreversible contributions which can arise in the first field
cycle, depending on the thermal history and field treatment of the
samples.
This can occur due to the existence of different meta-stable
multi-domain states, specifically for fields applied along $a$
and $\eta\rightarrow 0$.
  
\section{Conclusions and outlook}
\label{sec:outlook}

In summary, we have investigated the effect of epitaxial strain on the
caloric response of BTO through {\it{ab initio}}-based molecular
dynamics simulations. We have focused on tensile strain, which
provides a very promising route to enhance the ECE at room temperature
and below.  With increasing tensile strain, the ECE is systematically
reduced for fields along the surface normal. However, since the
optimal operation temperature is shifted towards ambient temperatures,
small values of tensile strain can be useful in order to optimize the
ECE at ambient temperatures. In addition, the ECE peak broadens with
increasing tensile strain, which is also favorable for applications.

For an electrical field applied within the surface plane, the ECE
exhibits two peaks, one slightly above the ferroelectric transition
temperature, $T_{\text{C}}$, and one close to the transition
temperature $T_{\text{S}}$ between the mono-domain and multi-domain
ferroelectric phases. Although, the latter is smaller than the peak at
$T_{\text{C}}$, it occurs in a temperature range that is very
attractive for many applications. Furthermore, $T_{\text{S}}$ can be
adjusted with the magnitude of the strain.
Between the strain-dependent temperatures $T_{\text{S}}$ and
$T_{\text{C}}$, we find a reversible coupling between external field
and the domain configuration. This strong coupling allows for a linear
increase of the ECE with the strength of the external field, and thus
large adiabatic temperature changes can be obtained within a broad
temperature range.
Thus, tensile epitaxial strain can be used to enhance the caloric
response of BTO well below the ferroelectric transition temperature,
and a considerable ECE around and below room temperature can be
achieved if the external electrical field is applied within the
surface plane.

Recently, it has been pointed out that the
electric field control of the domain structure in BTO can improve the
caloric response of FeRh films grown on BTO \cite{Liu}. Our results
suggest that the caloric response of the BTO domain structure itself
should also be taken into account, and that the resulting multicaloric
properties can be further optimized through epitaxial strain.
In the same spirit, one could also consider to use piezoelectric,
magnetostrictive, or mechanically-bendable substrates to adjust the
caloric response of BTO by shifting its optimal operation range to
higher and lower temperatures, which then allows to adjust the cooling
device to a specific  heat source.

\acknowledgements
We acknowledge financial support by the Deutsche
Forschungsgemeinschaft and the Swiss National Science Foundation (SPP
1599). Anna Gr\"unebohm thanks the Center for Computational Science
and Simulation (University of Duisburg-Essen) for computational time.


\begin{thebibliography}{0}
\bibitem{Planes}  A.~Planes,  L.~Ma{\~n}osa,  and  M.~Acet,   J. Phys.: Condens. Matter \textbf{21}, 233201 (2009).  
\bibitem{Moya} X.~Moya, S.~Kar-Narayan, and N.~D.~Mathur, Nat. Mater. {\textbf{22}}, 439, (2014).
\bibitem{Bai}  Y.~Bai, {G.-P.}~Zheng {K.~D.} Zheng, K.~Ding, L.~Qiao, S.~Q.~Shi, and D.~Guo,\  J. App. Phys.   {\textbf{110}},     {094103} ({2011}). 
\bibitem{Moya2} {X.}~{Moya},    {E.}~{Stern-Taulats}, {S.}~{Crossley}, {D.}~{Gonz\'{a}lez-Alonso}, {S.}~{Kar-Narayan},  
  {A.}~{Planes}, {L.}~\ {Ma{\~n}osa},  and {N.~D.~{Mathur}}, Adv. Mater.   {\textbf{25}}, {1360}  ({2013}).
\bibitem{ferrocooling}
S.~F\"ahler, U.~K.~R\"o{\ss}ler, O.~Kastner, J.~Eckert, G.~Eggeler, H.~Emmerich, P.~Entel, S.~M\"uller, E.~Quandt and K.~Albe
Adv. Eng. Mater., {\textbf{14}}, 10, (2012).
\bibitem{Valant} M.~Valant, Prog. Mater. Sci. {\textbf{57}}, 980 (2012).
\bibitem{Takeuchi/Sandeman:2015} I.~Takeuchi and K.~Sandeman, Physics Today {\bf 68}, 48 (2015).
\bibitem{Scott:2011} J.~F.~Scott, Annu. Rev. Mater. Res. {\bf 41}, 229 (2011).
\bibitem{Mischenko_et_al:2006} A.~S.~Mischenko,  Q.~Zhang,  J.~F.~Scott, R.~W.~Whatmore, R. W., and N.~D.~Mathur, Science {\bf{311}}, 1270 (2006).
\bibitem{Lisenkov} {S.}~{Lisenkov}  and {I.}~{Ponomareva}, Phys. Rev. B   {\textbf{80}},   {140102(R)} ({2009}).
\bibitem{Schlom} {D.~G.~{Schlom}}, {L.-Q.~{Chen}}, {C.-B.~{Eom}},
  {K.~M.~{Rabe}}, {S.~K.~{Streiffer}}, and {J.-M.~{Triscone}},
  Ann. Rev. Mater. Research {\textbf{37}}, {589} ({2007}).
\bibitem{pertsev2}  {N.~A.~{Pertsev}},      {A.~G.~{Zembilgotov}},  and   {A.~K.~{Tagantsev}}, Phys. Rev. Lett.   {\textbf{80}}, {1988} ({1998}).
\bibitem{pertsev4} N.~A.~{Pertsev}  and     {V.~G.~{Koukhar}}, Phys.  Rev. Lett.  {\textbf{84}},  {3722}  ({2000}).
\bibitem{pertsev3}  {N.~A.~{Pertsev}},      {V.~G.~{Koukhar}},      {R.~{Waser}},  and     {S.}~{Hoffmann}, Integrated  Ferroelectrics   {\textbf{32}},    {235}  ({2001}).
\bibitem{Dieguez} {O.}~{Di\'{e}guez}, {S.}~{Tinte}, {A.}~{Antons},
  {C.}~{Bungaro}, J.~B.~{Neaton}, K.~M.~{Rabe}, and {D.}~{Vanderbilt},
  Phys. Rev. B {\textbf{69}}, {212101} ({2004}).
\bibitem{Li2} Y.~L.~{Li}  and      {L.~Q.~{Chen}}, Appl. Phys.  Lett.  {\textbf{88}},\     {072905}  ({2006}).
\bibitem{BTO_strain} {A.}~{Gr\"unebohm},      {M.}~{Marathe}, and      {C.}~{Ederer}, Appl.  Phys. Lett.   {\textbf{107}},     {102901}  ({2015}).
\bibitem{Qiao} L.~Qiao and X.~Bi,  Appl. Phys. Lett. {\textbf{92}}, {062912} (2008).
\bibitem{Everhardt} A.~S.~Everhardt, S.~Matzen, N.~Domingo, G.~Catalan, and B.~Noheda, {Adv. Electron. Mater.}  {\textbf{2015}}, {1500214} (2015).
\bibitem{Akcay}{G.}~{Akcay},  {S.~P.~{Alpa}},  {G.~A.~{Rossetti}}, and {J.~F.~{Scott}}, J. Appl. Phys.   {\textbf{103}},  {024104} ({2008}).
\bibitem{Cao} {H.-X. ~{Cao}  and  {Z.-Y.~{Li}}, J. Appl. Phys.  {\textbf{106}},   {094104} ({2009})}.
\bibitem{Zhang3}{X.}~{Zhang},      J.~B.~Wang, {B.}~{Li}, X.~L.~Zhong,  {X.~J.~{Lou}}, and {Y.~C.~{Zhou}}, J. Appl. Phys. {\textbf{109}},  {126102} ({2011}).
\bibitem{Madhura}  M.~Marathe  and    {C.}~Ederer,\  App. Phys. Lett.   {\textbf{104}},     {212902} ({2014}).
\bibitem{Perantie} J.~Per\"antie, J.~Hagberg,  A.~Uusim\"aki, and H.~Jantunen,  Phys. Rev. B, {\textbf{82}}, 134119 (2010).
\bibitem{Ponomareva} I.~Ponomareva and S.~Lisenkov, Phys. Rev. Lett., {\textbf{108}}, 167604 (2012).
\bibitem{LeGoupil_et_al:2014} F.~Le Goupil, A.-K.~Axelsson,
  L.~J.~Dunne, M.~Valant, G.~Manos, T.~Lukasiewicz, J.~Dec,
  A.~Berenov, and N.~McN.~Alford, Adv. Energy Mater. {\bf 4}, 130688 (2014).
\bibitem{Zhong} {W.}~{Zhong},  {D.}~{Vanderbilt}, and  {K.~M.~{Rabe}}, Phys. Rev. B  {\textbf{52}},   {6301} ({1995}).
\bibitem{Feram1}{T.}~{Nishimatsu},      {U.~V.~{Waghmare}},   {Y.}~{Kawazoe},  and     {D.}~{Vanderbilt}, Phys. Rev. B {\textbf{78}}, {104104} ({2008}).
\bibitem{Nishimatsu} {T.}~{Nishimatsu},     {M.}~{Iwamoto},     {Y.}~{Kawazoe},  and      {U.~V.~{Waghmare}}, Phys Rev B  {\textbf{82}},    {134106} ({2010}).
\bibitem{Nose} {S.~D.~{Bond}},      {B.~J.~{Leimkuhler}},  and      {B.~B.~{Laird}}, Journal of Computational Physics {\textbf{151}},     {114 } ({1999}).  %
\bibitem{Marathe} {M.}~{Marathe}, {A.}~{Gr\"unebohm},
  {T.}~{Nishimatsu}, {P.}~{Entel}, and {C.}~{Ederer}, Phys. Rev. B
  {\textbf{93}}, {054110} ({2016}).
\bibitem{Nishimatsu3} T.~Nishimatsu, J.~A.~Barr, and S.~P.~Beckman,
  J. Phys. Soc. Jpn. {\textbf{82}},\ {114605} ( {2013}).
\bibitem{Liu2} Y.~Liu, J.~F.~Scott, and B.~Dkhil,
  Appl. Phys. Rev. {\textbf{3}},\ {031102} ( {2016}).  
\bibitem{Defects} A.~{Gr{\"u}nebohm} and {T.}~{Nishimatsu},
  Phys. Rev. B. {\bf{93}}, 134101 (2016).
  \bibitem{Liu}  Y.~Liu, L.~C.~Phillips, R.~Mattana, M.~Bibes, A.~Barth{\'e}l{\'e}my, and B.~Dkhil, Nat. Commun. {\textbf{7}}, 11614 (2016).
\end{thebibliography}
\end{document}